\documentstyle[prl,epsf,aps]{revtex}

\draft
\title{Fractional transport equations for L\'evy stable processes} 
\author{Eric Lutz} 
\address{Max--Planck--Institut f\"ur Kernphysik, Postfach 103980,
69029 Heidelberg, Germany \footnote{Present address : D\'epartement de
  Physique Th\'eorique, 
Universit\'e de Gen\`eve
24, quai Ernest Ansermet
CH-1211 Gen\`eve 4}}  
\date{\today}

\newcommand{\la}{\langle}
\newcommand{\ra}{\rangle}

\newcommand{\be}{\begin{equation}}
\newcommand{\ee}{\end{equation}}
\newcommand{\bea}{\begin{eqnarray}}
\newcommand{\eea}{\end{eqnarray}}

\begin{document}

\maketitle

\begin{abstract}
The  influence functional method of Feynman and Vernon is used to obtain   a
quantum  master equation 
for a Brownian system subjected to a  L\'evy stable  random force. The
corresponding classical transport equations for the Wigner function
are then derived,  both in the limit of weak and strong friction. These  are
fractional extensions of  the Klein--Kramers and the 
 Smoluchowski equations. It
is shown that the fractional character acquired by the  {\it position } in the
Smoluchowski equation follows from  the fractional character  of
the {\it momentum} in the Klein--Kramers equation. Connections among
 fractional transport equations recently
proposed are clarified. 
\end{abstract}
\vskip1pc

In the theory of Brownian motion one is interested in the time
evolution of  a system 
coupled to  a large environment. The effect of the
coupling is modeled by a stochastic force $F(t)$ with a given
probability density 
$P[F(t)]$. The dynamics of a Brownian particle of mass $M$
in the presence  of  an external potential $U(x)$ 
is  then  described by  the Langevin equation
\be 
\label{eq1}
M\ddot x(t) + \gamma \dot x(t) + U'(x) = \xi(t)\ ,
\ee
where $F(t)$  has been divided into a mean force proportional to the
velocity, the friction force  $-\gamma \dot x(t)$, plus  a fluctuating
part $\xi(t)$. In the usual treatment of Brownian motion \cite{gar90},
it is assumed  that the random force 
is Gaussian distributed with variance $\la \xi(t) \xi(t')\ra =
2D\,\delta(t-t')$ 
where $D=\gamma kT $ is the diffusion coefficient, and the Langevin
equation is shown to be fully equivalent to a phase--space equation |
the Klein--Kramers equation. In the limit of 
strong friction, the inertial term in the Langevin equation  can be
neglected and  the Klein--Kramers equation reduces to the Smoluchowski
equation. However, it has become clear in
recent years that many processes in nature, such as anomalous
diffusion (for a review see \cite{bou90,shl94,pek98}), cannot be
described by ordinary (Gaussian) Brownian motion. A case in point is
the so--called L\'evy flight  with a  stochastic
force  distributed according to  L\'evy stable statistics and  which  has
been introduced in connection with 
super--diffusion \cite{fog94,jes99}. In this Letter we consider the
generalization of
transport equations to  describe  L\'evy Brownian motion. This
question
 has already been addressed in the past  by using various methods
\cite{fog94,jes99,all96,yan00}, in particular the CTRW formalism
\cite{com96,met99}. However,  all these approaches were limited  to
coordinate space only. Here we present the first derivation of the
Klein--Kramers equation for a L\'evy stable process. We consider both
the case of a symmetric and asymmetric probability distribution. As our main tool, we  employ the influence functional
formalism developed by Feynman and Vernon \cite{fey63,fey65}.
     
If initially system and  environment are
 not correlated then,  according to Feynman and Vernon \cite{fey63,fey65}, the density
 operator of the system at time $t$ can be written in coordinate
 representation as
\be
\label{eq10}
\rho(x,x',t) = \int {\cal F}[x,x']\, \exp \frac{i}{\hbar} \Big
  \{S[x]-S[x']\Big \}\, \rho(x_0,{x_0}',0)\,\, {\cal D}x(t)
{\cal D}x'(t) dx_0 d{x_0}' \ .
\ee
Here the entire information on the coupling to the environment is
contained in the influence functional  ${\cal F}[x,x'] =
\Phi[\frac{x(t)-x'(t)}{\hbar}\Big]$, where $\Phi[k(t)]$
is the characteristic functional of the probability density P[F(t)],
\be
\label{eq11}
  \Phi[k(t)] = \int \exp\Big\{ i
\int F(t) k(t)dt\Big\} \,\, P[F(t)]{\cal D}F(t) \ . 
\ee      
If $F(t)$ is Gaussian distributed with mean $\la F(t)\ra = \gamma(t)$
and variance
$\la [F(t)-\gamma(t)][F(t')-\gamma(t')]\ra = D(t-t')$, the characteristic functional is
given by
\be
\label{eq12}
\Phi_G[k(t)] = \exp\Big\{i\int \gamma(t) k(t) dt - \frac{1}{2}\int \int
D(t-t') k(t) k(t') dtdt'\Big\}\ . 
\ee
If we further assume that the friction force  is proportional to the velocity of the 
system, $\gamma(t) = -\gamma [\dot x(t)+ \dot x'(t)]/2$,  and that the
variance  is delta correlated in time, 
$D(t-t') = 2D\, \delta(t-t')$, then the influence functional can be written in
the form
\be
\label{eq13}
{\cal F}[x,x'] = \exp \frac{i}{\hbar} \int \Big\{-\frac{\gamma}{2}\,
[x(t)-x'(t)][\dot x(t) + \dot x'(t)] + i\frac{D}{\hbar}\,[x(t)-x'(t)]^2
\Big\}dt \ .
\ee
By means of  a small time expansion, 
Eq.~(\ref{eq10}) can be transformed into a
differential equation for the density operator. Using the influence
functional (\ref{eq13}), this results in  the master equation
\be
\label{eq14}
i\hbar \frac{\partial\rho(x,x',t)}{\partial t} = \Big[H(x)-H(x') +
\frac{\gamma}{2M}\,(x-x')(p_x-p_{x'})  - i\frac{D}{\hbar}\,(x-x')^2\Big] \rho(x,x',t)\ ,
\ee
where $H(x) = {p_x}^2/2M + U(x)$ is the Hamiltonian of the system. We
recognize in Eq.~(\ref{eq14}) the master equation for quantum Brownian 
motion derived by Caldeira
and Leggett using the oscillator bath model \cite{cal83} (see also
\cite{lut99}). 
  
In the case of a  L\'evy stable distribution, the
characteristic functional is given by \cite{gne54}
\be
\label{eq15}
\Phi_{L}[k(t)] = \exp\Big\{i\int \gamma(t)k(t) dt - \int
C(t) \,|k(t)|^\alpha \Big[1+i\beta \frac{k}{|k|}\tan\frac{\alpha\pi}{2}\Big] dt
\Big\}\ , 
\ee
where $\alpha$ ($0<\alpha\leq 2$) is the characteristic exponent (or
stability index) of 
the distribution  and
$\beta$ ($-1\leq \beta \leq 1)$ is the asymmetry parameter \cite{not}. 
We assume as before that $\gamma(t)$ is proportional to the
velocity of the system and take $C(t) = D$ constant. This leads to the 
following quantum master equation for a L\'evy stable process
\be
\label{eq16}
i\hbar \frac{\partial\rho(x,x',t)}{\partial t} = \Big[H(x)-H(x')+
\frac{\gamma}{2M}\,(x-x')(p_x-p_{x'})  -
i\frac{D}{\hbar^{\alpha-1}}\,|x-x'|^{\alpha-1} \nonumber \Big(|x-x'| +i\beta
\tan\frac{\alpha\pi}{2} (x-x')\Big)\Big] \rho(x,x',t)\ .
\ee
For $\alpha = 2$ the master equation (\ref{eq16}) reduces to the
Caldeira--Leggett  equation (\ref{eq14}).
In order to obtain the corresponding classical transport equation, we
introduce the 
Wigner transform of the density matrix 
\be
\label{eq17}
f(q,p,t)= \frac{1}{2\pi\hbar}\int_{-\infty}^\infty dr \exp\Big[-\frac{ipr}{\hbar}\Big]
\rho\Big(q+\frac{r}{2}, q-\frac{r}{2},t\Big)  \ .
\ee
Applying the Wigner transform to Eq.~(\ref{eq16}) and keeping only terms in
leading order in $\hbar$, we obtain the 
equation
\be  
\label{eq18}
\frac{\partial f}{\partial t}= -\frac{p}{M}\frac{\partial f}{\partial
  q}+ U'(q)\frac{\partial f}{\partial p}+\frac{\gamma}{M}
 \frac{\partial }{\partial p}(pf) + \gamma kT \left [\frac{\partial^\alpha
  f}{\partial |p|^\alpha}+\beta
 \tan\frac{\alpha\pi}{2}\,\frac{\partial}{\partial p}\frac{\partial^{\alpha-1}
  f}{\partial |p|^{\alpha-1}}\right ]\ ,
 \ee
where we have introduced the Riesz fractional derivative which is defined
through its Fourier transform as \cite{sam93,sai97} 
\be 
\label{eq19}
 -\frac{\partial^\alpha}{\partial
  |p|^\alpha} = \frac{1}{2\pi\hbar}\int_{-\infty}^\infty dr\,
\exp\Big[-\frac{ipr}{\hbar}\Big]\, \frac{|r|^\alpha}{\hbar^\alpha}\ .
\ee
The fractional equation (\ref{eq18}) for the distribution function $f(q,p,t)$
describes the complete dynamics of the Brownian system  in
phase--space,  for  both symmetric and asymmetric L\'evy stochastic
forces. In the latter case,  we observe an
additional contribution to the friction term which may be of relevance 
for the description of 
anomalous transport in anisotropic media \cite{cha98}. It is worthwhile to note that the fractional character in
Eq.~(\ref{eq18}) is carried by the {\it momentum}.  For $\alpha = 2$ we recover the ordinary 
Klein--Kramers equation. In the limit of high friction, one may
exploit the rapid relaxation of the momentum
distribution  to a stationary distribution, to write down 
 a simplified equation for the reduced distribution function in
 configuration space, $\hat f(q,t)= \int dp f(q,p,t)$. Employing  the
 systematic expansion method developed in Ref.~\cite{tit78}, we obtain 
 the following fractional extension of the Smoluchowski equation (the
 derivation of Eq.~(\ref{eq20}) will be sketched below)
\be 
\label{eq20}
\frac{\partial \hat f}{\partial t}=
\frac{1}{\gamma}\frac{\partial}{\partial q}(U'(q)\hat f) + \frac{kT}{\gamma} \Big[\frac{\partial^\alpha
 \hat  f}{\partial |q|^\alpha} +\beta \tan\frac{\alpha\pi}{2} \frac{\partial}{\partial q}\frac{\partial^{\alpha-1}
  \hat f}{\partial |q|^{\alpha-1}}\Big ]\ . 
\ee
We  see that the fractional character has  been transferred to the
{\it position}. This observation settles an apparent point of
confusion in the literature. Indeed, in most of the phenomenological
fractional 
generalizations of transport equations to phase--space, it is  unclear
whether the 
fractional derivative should be taken with respect to position or
momentum or even both (see e.g. the discussion in
Refs.~\cite{kus99,bar00}). Our findings show that for a L\'evy flight
the fractional character 
acquired by the position in the Smoluchowski equation follows from the 
fractional character of the momentum in the Klein--Kramers
equation. 

Let us now discuss the connections of the  fractional 
Smoluchowski equation (\ref{eq20}) to the equations considered in the
literature. We first begin with the case of a symmetric L\'evy force. Equation (\ref{eq20}) with $\beta = 0$
(which in this form is also sometimes called fractional Fokker--Planck
equation)   has 
been studied in  detail in Ref.~\cite{jes99} for  the cases of a free
flight, a particle subjected to a  constant force  and to 
a linear Hookean force. Moreover, it is interesting to note  that
Eq.~(\ref{eq20}) has been recently derived in 
Ref.~\cite{met99} from a
generalized master equation for  a non--homogeneous random walk. The
corresponding fractional diffusion 
equation obtained by setting $U(q)=0$  has also been
considered in Ref.~\cite{sai97}. On the other hand, for an asymmetric
random force,  $\beta \neq 0$, 
 the Smoluchowski equation (\ref{eq20}) is a generalization  to a
 velocity dependent 
damping force of a fractional diffusion equation recently obtained  in
Ref.~\cite{yan00} starting
from a Langevin--like equation.  

In a recent work, Kusnezov {\it et al.} \cite{kus99} proposed a
fractional Klein--Kramers equation which was obtained as the classical
limit of what they call a quantum L\'evy 
process. Their derivation is based on  a microscopic random--matrix model for a
system coupled to a chaotic environment. These authors showed that for
the case of an 
environment with constant
average level density  (or equivalently  with infinite temperature $\beta_T = (kT)^{-1} = 0$), the reduced density
matrix  displays the behaviour of 
a free L\'evy flight. However, the dynamics described by their
fractional transport equation for finite temperature $\beta_T \neq 0$
is unknown. Let us now examine that point.     
The characteristic functional corresponding to their quantum master
equation can be written as
\be
\label{eq21}
\Phi_{QL}[k(t)] = \exp\Big\{i\int \gamma(t) \, \frac{\alpha}{2} \mbox{ sgn} k(t)\, |k(t)|^{\alpha-1}  dt -  \int
C(t)\, |k(t)|^\alpha  dt
\Big\}\ . 
\ee
By comparing  expression (\ref{eq21}) to the characteristic functional of a
symmetric L\'evy flight (\ref{eq15}), we
observe (i) that the second terms on the rhs, those   describing   the
fluctuation properties 
of the stochastic force, are equal, but (ii) that the first terms,  which 
are  related to 
the mean, are  different. Since the latter are  responsible for the
dissipation, this implies  that two expressions    are
identical only for vanishing friction. Note that this is in  agreement 
with the results of Ref.~\cite{kus99}, since the limit of vanishing friction 
precisely
corresponds to $ \beta_T=0$,
 as can be easily seen from their equation (28). The
fractional Kramers equation (27) (with $\beta_T\neq0$) given in
Ref.~\cite{kus99}  
thus describes a Brownian system subjected to a symmetric
L\'evy stable  random 
force, but with a  mean friction force that is {\it different} from that of a
L\'evy flight.  
 It is straightforward to determine the mean value  $\la F_{QL}(t)
 \ra$ for the process defined by Eq.~(\ref{eq21}). It is given at a
 particulary 
time $t=t_0$ by \cite{fey65}
\be
\label{eq22}
\la F_{QL}(t_0)\ra = -i \, \left. \frac{\delta \Phi_{QL}[k(t)]}{\delta
    k(t_0)}\right|_{k=0} \ .
\ee
Since $\alpha-1 \leq 1$, the mean force $\la F_{QL}(t_0)\ra$
is divergent (it is finite and equal to $\gamma(t_0)$ only for
$\alpha=2$). 
Hence both the first and the second moment of the process
investigated by Kusnezov {\it et al.} are divergent. This has to  
be contrasted with the normal L\'evy flight, where the mean is finite
and equal to $\la F_L(t_0)\ra =
\gamma(t_0)$ for all values of $\alpha$. Let us emphasize 
that  the random--matrix results are  therefore at
variance with the continuous time random walk calculation \cite{met99}. The two approaches only agree for the special
case of a Gaussian distribution $\alpha=2$. The divergence of the
damping force in the random--matrix model can be seen  to be  related to the Markovian
property of the transport equation.

We now return to the derivation of Eq.~(\ref{eq20}). For simplicity we 
will only consider the case of a symmetric probability distribution. For large $\gamma$, the
dynamics of  the Klein--Kramers  equation (\ref{eq18}) is dominated by
the term which  contains the operator 
\be
\label{eq25}
C = \frac{\partial}{\partial p}p + MkT
\frac{\partial^\alpha}{\partial|p|^\alpha} \ .
\ee
We shall look for an approximation  to leading
 order in $(\gamma/M)^{-1}$ of  Eq.~(\ref{eq18}) by using an eigenvalue method \cite{tit78}.
We denote by $\varphi_n(p)$ the eigenfunctions of the operator $C$ and by
$-n \,(n= 0, 1, 2, ...)$ the corresponding eigenvalues. We  define the
following raising and lowering operators
\be
\label{eq26}
a_+ = -kT \frac{\partial}{\partial p} \hspace{1cm}\mbox{ and
  }\hspace{1cm} a_- = M D^\alpha_p + 
\frac{1}{kT} \, p \ ,
\ee
where the operator $D^\alpha_p$ obeys $\frac{\partial}{\partial p}\,
D^\alpha_p =\frac{\partial^\alpha}{\partial|p|^\alpha}$. The two
operators $a_+$ and $a_-$ 
satisfy $ C= -a_+a_-$ and $[a_-,a_+] =1$ and we have  further the
ladder relations $a_+\varphi_n(p)=
(n+1)\,\varphi_{n+1}(p)$ and $a_-\varphi_n(p)=(1-\delta_{n,0})\,\varphi_{n-1}(p)$.
Next we  look for a  solution of Eq.~(\ref{eq18})
in the form
\be 
\label{eq27}
f(q,p,t) = \hat f(q,t)\, \varphi_0(p) + \left[\frac{\gamma}{M}\right]^{-1}\, f^{(1)}(q,p,t)
+ \left[\frac{\gamma}{M}\right]^{-2}\,f^{(2)}(q,p,t)+ ...
\ee
and
\be 
\label{eq28}
\frac{\partial \hat f(q,t)}{\partial t} = \left( \partial^{(0)} +
\left[\frac{\gamma}{M}\right]^{-1}\, \partial^{(1)}+\left[\frac{\gamma}{M}\right]^{-2}\, \partial^{(2)}+ 
...\right) \hat f(q,t)\ ,
\ee
where the $ \partial^{(i)}$ are linear differential operators which
are determined as follows: we substitute Eqs.~(\ref{eq27}) and
(\ref{eq28}) into Eq.~(\ref{eq18}) and separate the different orders
in $(\gamma/M)^{-1}$.  The integrability condition then yields for the
two lowest orders 
 
\be
\partial^{(0)} = 0 \hspace{1cm}\mbox{ and } \hspace{1cm}\partial^{(1)}
= \frac{kT}{M}\frac{\partial}{\partial q}\left[D^\alpha_q +
  \frac{1}{kT}\, U'(q)\right ]\ .
\ee
The operator $\partial^{(1)}$ is precisely the one appearing in the
Smoluchowski equation (\ref{eq20}).  

To conclude, using  influence functional methods,  we gave
the first derivation of a 
quantum master equation  for symmetric and asymmetric
L\'evy flights with viscous damping. By taking  the classical
limit, we then obtained an extension of the
Klein--Kramers equation
containing fractional derivatives with respect to {\it momentum}. In the
limit of strong damping, this equation was shown to reduce to a
fractional Smoluchowski equation with  fractional
derivatives with respect to {\it position}. Furthermore, for
symmetric L\'evy stable laws,
our results 
are in agreement with those of Kusnezov {\it et al.} in the limit of vanishingly
small friction. For non--zero friction, we found that  the process 
described by their fractional Kramers equation possesses a divergent
mean damping force. In the opposite limit of strong 
friction, we recovered the fractional Fokker--Planck equations
considered in Refs.~\cite{jes99,met99}. Finally, for the case of
asymmetric L\'evy  
stable distributions, we gave an extension to a velocity dependent
damping force  of a fractional diffusion 
equation suggested in Ref.~\cite{yan00}.


\begin{references}
\bibitem{not} For simplicity we only consider the case $\alpha \neq1$, 
  or $\beta = 0$.
\end{references}
\end{document}